\documentclass[11pt]{article}
\usepackage{graphicx} 
\usepackage{amsmath,amssymb, mathtools,siunitx, multicol, hyperref}
\usepackage[left=2cm, right=2cm, bottom=1in, top=1in]{geometry}
\usepackage[backend=biber,style=numeric,sorting=none,maxbibnames=3,     
    giveninits=true]{biblatex}
\DeclareNameAlias{default}{family-given}
\DeclareNameAlias{author}{family-given}
\DeclareNameAlias{given}{initials}

\newcommand{\apjl}{Astrophys. J. Lett.\ }

\newcommand{\apjs}{Astrophys. J. Suppl. Ser.\ }
\newcommand{\aap}{Astron. Astrophys.\ }

\AtEveryBibitem{%
  \clearfield{title}%
  \clearfield{month}%
  \clearfield{day}%
  \clearfield{doi}%
  \clearfield{issn}%
  \clearfield{isbn}%
  \clearfield{note}%
  \clearname{editor}%
}
\renewbibmacro*{in:}{}

\DeclareFieldFormat{eprint}{\href{https://arxiv.org/abs/#1}{\textcolor{myblue}{arXiv}}}

\renewbibmacro*{eprint+url}{%
  \iffieldundef{eprint}{}{\printfield{eprint}}%
} 

\DeclareFieldFormat{url}{\href{#1}{\textcolor{blue}{arXiv}}}

\renewbibmacro*{url+urldate}{%
  \iffieldundef{url}{}{%
    \printfield{url}%
  }%
}

\usepackage{authblk}
\usepackage{siunitx}
\usepackage{tcolorbox}
\usepackage{titlesec}
\titlespacing*{\section}
{0pt}{1.5ex plus 1ex minus .1ex}{0.5ex plus .1ex}

\titleformat{\section}
    {\normalfont\large\bfseries}{\thesection}{1em}{}

\newcommand{\daa}{\ensuremath{\Delta\alpha/\alpha}}

\usepackage{aas_macros}

\title{Beyond $\Lambda$CDM: fundamental constants as cosmological observables}

\author[1,2]{Dinko Milakovi{\'c}\thanks{Corresponding author: \href{mailto:dinko@milakovic.net}{dinko@milakovic.net}}}
\author[3,4]{John Webb}

\affil[1]{INAF-OATs, Trieste Observatory, Via Tiepolo 11, I-34131 Trieste, Italy}
\affil[2]{Institute for Fundamental Physics of the Universe (IFPU), via Beirut 2, I-34151 Trieste, Italy}
\affil[3]{Institute of Astronomy, Madingley Rd, Cambridge CB3 0HA, United Kingdom}
\affil[4]{Department of Physics, Beijing Normal University, Zhuhai 519087, China}

\begin{document}

\maketitle

\newpage
\section{Why fundamental constants matter in cosmology}

Precision cosmology has entered an era in which gravity is exquisitely measured, but the laws of physics themselves remain largely untested on cosmic scales. Over the coming decade, a new generation of flagship facilities, including the Vera C. Rubin Observatory, \emph{Euclid}, the Dark Energy Spectroscopic Instrument (DESI), the Square Kilometre Array, and next-generation cosmic microwave background (CMB) experiments, will deliver extraordinarily precise measurements of cosmic structure and expansion. Together, these surveys will map the Universe through the distance–redshift relation, baryon acoustic oscillations, the growth of large-scale structure, and weak and strong gravitational lensing. Despite their breadth and power, however, all of these probes share a fundamental limitation: they interrogate the dark sector exclusively through gravity. If the physics underlying dark energy or dark matter involves new degrees of freedom that couple only weakly, indirectly, or non-universally to baryonic matter, then even the most exquisite gravitational datasets may provide only an incomplete view of the underlying physics.

Searches for spacetime variations in the fine-structure constant,
$\alpha = {e^2}/{4\pi\epsilon_0\hbar c}$,
offer a qualitatively different and highly complementary approach \cite{Uzan2025LRR....28....6U}. In any consistent relativistic framework, variations in a dimensionless constant such as $\alpha$ necessarily imply the existence of new dynamical degrees of freedom, typically light scalar fields, that couple to electromagnetism and may also interact with the dark sector. Measurements of $\alpha$ therefore probe fundamental physics directly, rather than indirectly through gravitational dynamics, and are uniquely sensitive to new interactions that may leave only subtle or degenerate signatures in large-scale structure.

In a broad class of such theories, spatial perturbations in $\alpha$ evolve dynamically during the matter-dominated era, with unscreened sub-horizon modes growing approximately as $\delta\alpha/\alpha \propto (1+z)^{-1}$ \cite{Mota2004MNRAS}.
Consequently, even extremely small primordial fluctuations present at recombination can be significantly amplified by late cosmic times. Moving backwards in time, a constraint on spatial fluctuations in $\alpha$ at a precision of, say, $z=3$ of $\daa \sim 10^{-6}$ therefore corresponds to an impressive constraint of $\sim 10^{-9}$ at the recombination epoch (a far more stringent constraint than temperature fluctuations from the CMB data itself). 

A natural next step is therefore to sample spacetime with a high surface density of precision $\alpha$ measurements and to cross-correlate these with tracers of the matter distribution and with CMB-derived maps of gravitational potential. This approach opens a discovery space inaccessible to existing and planned facilities. By mapping $\alpha$ across space and time, we can test whether the laws of physics themselves respond to cosmic environment, large-scale structure, or horizon-scale initial conditions; such signatures arise naturally in theories involving new quantum fields, grand unification, or screened fifth forces.

The scientific case is further strengthened by mounting observational tensions and anomalies suggesting possible departures from the $\Lambda$CDM paradigm. In this context, a large-scale, high-precision survey of $\alpha$ is both timely and essential: it provides a fundamentally different diagnostic of new physics at a moment when cosmology increasingly hints that such physics may be required. Even null results would be transformative, placing the most stringent constraints on the coupling between electromagnetism and the dark sector, restricting entire classes of models beyond $\Lambda$CDM and dark energy theories. In this sense, a large-scale $\alpha$-variation survey is highly complementary to flagship cosmological programmes, since it probes the dark sector through the values of fundamental constants rather than through spacetime geometry alone.

\begin{tcolorbox}[
    colback=white, 
    colframe=blue, 
    boxsep=1pt, 
    left=5pt,          
    right=5pt,
    top=2pt,
    bottom=2pt,
    fonttitle=\bfseries
]
    While upcoming cosmological surveys will map the Universe with extraordinary precision, measurements of fundamental constants test something equally profound: whether the laws of physics themselves are dynamical on cosmological scales.
\end{tcolorbox}

\section{Spectroscopic signatures of varying $\alpha$}
Any variation in the value of $\alpha$ will perturb the energy levels of atoms, shifting the observed transition frequencies away from their laboratory values: $\omega_{obs} \approx \omega_{lab} +2 q \daa$, where $q$ is a theoretical sensitivity coefficient and $\Delta\alpha/\alpha \equiv (\alpha_{obs}-\alpha_{lab})/\alpha_{lab}$ \cite{Dzuba1999}. Two spectroscopic methods have been used to map cosmological $\alpha$ variations: the ``alkali doublet'' (AD) \cite{Savedoff1956Natur.178..688S} and the ``many multiplet'' (MM) method \cite{Dzuba1999,Webb1999}. 

The AD method utilizes the distance between fine-structure doublets of forbidden emission lines in galaxies, such as the [O III] $\lambda\lambda$ 4959, 5007, which scales as $\alpha^2$. Because emission line galaxies (ELGs) have very broad lines ($\mathrm{FWHM}\sim\qty{100}{\kilo\meter\per\second}$) and only one doublet is usually available for measurement, constraints on $\daa$ are $\sim10^{-4}$. The most precise constraints come from applying the MM method to quasar absorption systems. As the light from a background quasar passes through cool gas clouds (damped Lyman-$\alpha$ systems or Lyman-limit systems), it imprints metal absorption lines (e.g., Mg, Fe, Zn, Cr). Each transition has a different $q$ coefficient, so $\daa$ can be constrained from relative velocity shifts between transitions. The larger number of available transitions and their narrow intrinsic widths ($\mathrm{FWHM}\lesssim\qty{10}{\kilo\meter\per\second}$) allow constraining $\daa$ with a precision of $\sim10^{-6}$ in a single system.

\section{Measuring $\alpha$ in the 2040s}
The coming decades offer a unique opportunity to transform searches for variations in the $\alpha$ from a niche endeavour into a precision, survey-driven branch of fundamental cosmology. The Wide-field Spectroscopic Telescope (WST) is a proposed new facility with a 12-meter radius mirror, planned for the 2040s \cite{Mainieri2024arXiv240305398M}. The observational capabilities of WST, combined with methodological advancements, will open new opportunities for the field.

First, wide-field high-resolution spectroscopy on facilities such as the WST will enable $\sim 100,000$ new MM measurements at a spectral resolution of $R \approx 40,000$. This represents an increase of roughly two orders of magnitude over the total number of high-resolution quasar absorption measurements available in 2025. Crucially, these measurements will not only be more numerous but will also be obtained at substantially higher signal-to-noise ratios (than existing UVES measurements), leading to markedly smaller statistical uncertainties and improved control of instrumental systematics. Prior to the advent of WST, there is little prospect that this situation will change dramatically; although ESPRESSO and ANDES will make important contributions of a small number of measurements at unprecedented spectral resolution, only a wide-field fibre-fed facility such as WST have the capability to make such a dramatic advance in terms of sample size, survey speed, and sky coverage at high redshift.

Second, WST will extend this advance by enabling an unprecedented 1 million new AD measurements across the sky, using the [O III] $\lambda\lambda$ 4959, 5007, and [Ne III] $\lambda \lambda$ 3869, 3967 lines of galaxies at redshifts $z<1$ ($R\approx4,000)$. This dwarfs the current state of the art, where approximately 110,000 measurements have been obtained using DESI \cite{Jiang2024_DESI}. While individual AD constraints are intrinsically less sensitive than MM measurements, their sheer number and uniform selection will allow $\alpha$ to be mapped statistically across cosmic time and large angular scales; in combination with high-redshift measurements, this will enable powerful consistency checks, cross-calibration between methods, and the first genuinely tomographic studies of $\alpha$ as a cosmological field.

Third, and perhaps most important, recent detailed analyses using sophisticated forward-modelling techniques, notably AI-VPFIT \cite{Lee2021aivpfit}, have demonstrated that the majority of existing high-resolution measurements are strongly biased towards a null result\cite{Webb2022}. These biases arise from a combination of human-guided model selection, limited exploration of complex parameter spaces, and subtle instrumental and astrophysical systematics. As a result, the current empirical status of varying-$\alpha$ measurements is ill-determined: the absence of a convincing detection cannot be straightforwardly interpreted as evidence that variations do not exist. It remains entirely plausible that genuine signals have been systematically suppressed or misidentified.

\section{Technology and data handling requirements}
Making high quality measurements requires excellent control over the wavelength calibration, either using astronomical laser frequency combs (LFC) or Fabry-P{\'e}rot etalons to remove any wavelength distortions that may emulate or hide a varying $\alpha$ signal. A further requirement would be to rigorously establish the spectroscopic instrumental profile across the entire field of view, for optimal absorption line centroid accuracy. Finally, the new, statistically powerful datasets must be analysed using fully automated and reproducible techniques \cite{Lee2021aivpfit}, to eliminate previous human bias, and for feasibility (interactive modelling would be impossible for such a large database). Only in recent years has the community begun to understand the dominant sources of bias in earlier work, and by the time WST becomes operational, robust analysis frameworks will be firmly established. 


\begin{figure}
    \centering
    \includegraphics[width=0.7\linewidth]{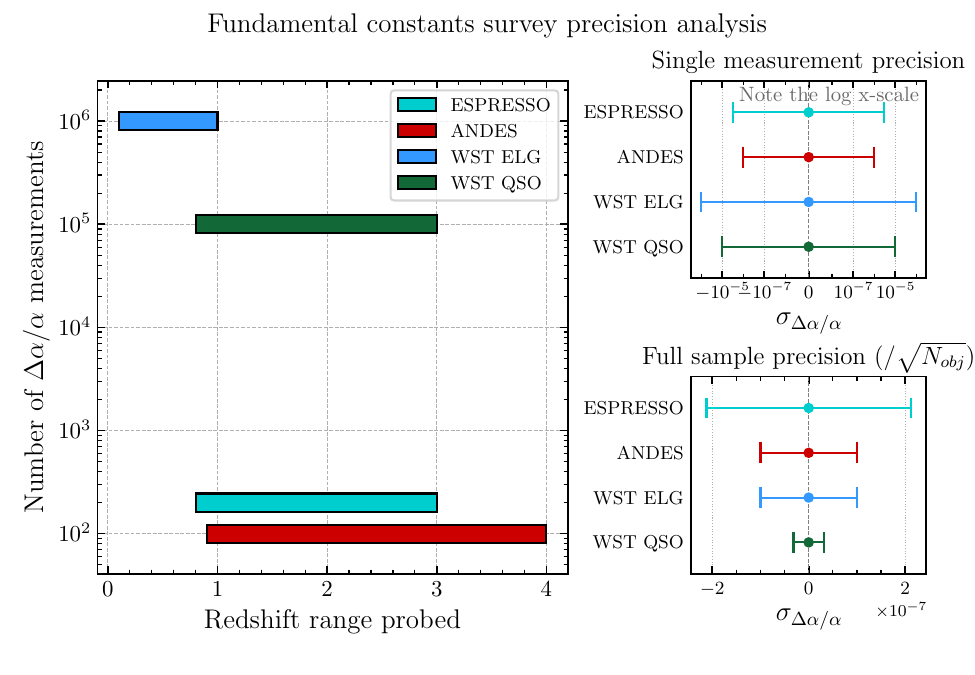}
    \vspace{-2em}
    \caption{$\alpha$ measurements in the 2040s. Left: a comparison of the number of {\daa} measurements from two single-object spectrographs (ESPRESSO and ANDES), as well as the WST programmes targeting emission line galaxies (ELG) and quasars (QSO) in this proposal. Instrumental systematics limit DESI measurements to $|\daa|<10^{-4}$, so it is not shown. The WST ELGs are the only ones covering $z<1$ and the WST QSO sample will represent an unprecedented jump in the number of measurements. Top right: the expected statistical uncertainty on $\daa$ for a single measurement. Bottom right: the final statistical uncertainty on {\daa} from the full measurement sample. }
    \label{fig:alpha}
\end{figure}

\printbibheading
\begin{multicols}{2}
\printbibliography[heading=none]
\end{multicols}
\end{document}